\documentclass[conference]{IEEEtran}
\IEEEoverridecommandlockouts

\usepackage[utf8]{inputenc} 
\usepackage[T1]{fontenc}    
\usepackage{listings}       

\usepackage{iftex}
\ifpdf
\usepackage{underscore} 
\usepackage[T1]{fontenc} 
\else
\usepackage{breakurl}
\fi

\usepackage{amsmath}
\usepackage{graphicx}
\usepackage{float}
\graphicspath{{Figures/}} 

\usepackage{listings}
\usepackage{xcolor}
\usepackage{cite}
\usepackage{tikz}
\usetikzlibrary{positioning,fit,arrows.meta}
\usepackage{rotating}
\usepackage{multirow}

\usepackage{etoolbox}
\usepackage{balance}

\title{Towards Continuous Assurance with Formal Verification and Assurance Cases}
\author{
	Dhaminda B. Abeywickrama$^{*1}$ \quad
	Michael Fisher$^{1}$ \quad
	Frederic Wheeler$^{2}$ \quad
	Louise Dennis$^{1}$\\
	\\
	$^{1}$Department of Computer Science, The University of Manchester, Manchester, UK\\
	\texttt{\{dhaminda.abeywickrama, michael.fisher, louise.dennis\}@manchester.ac.uk}\\
	\\
	$^{2}$Regulatory Support Directorate, Amentum, Warrington, UK\\
	\texttt{frederic.wheeler@global.amentum.com}
}

\begin{document}
	\maketitle
	
	\begin{abstract}
		Autonomous systems must sustain justified confidence in their correctness and safety across their operational lifecycle---from design and deployment through post-deployment evolution. 
		Traditional assurance methods often separate development-time assurance from runtime assurance, yielding fragmented arguments that cannot adapt to runtime changes or system updates --- a significant challenge for \emph{assured autonomy}. Towards addressing this, we propose a unified \emph{Continuous Assurance Framework} that integrates design-time, runtime, and evolution-time assurance within a traceable, model-driven workflow as a step towards \emph{assured autonomy}. In this paper, we specifically instantiate the design-time phase of the framework using two formal verification methods: RoboChart for functional correctness and PRISM for probabilistic risk analysis. 
		We also propose a model-driven transformation pipeline, implemented as an Eclipse plugin, that automatically regenerates structured assurance arguments whenever formal \emph{specifications} or their \emph{verification} results change, thereby ensuring traceability. 
		We demonstrate our approach on a nuclear inspection robot scenario, and discuss its alignment with the \emph{Trilateral AI Principles}, reflecting regulator-endorsed best practices.
	\end{abstract}
	
	\begin{IEEEkeywords}
		continuous assurance, dynamic assurance, formal verification.	
	\end{IEEEkeywords}
	
	\section{Introduction}
	\label{sec:introduction}
	Autonomous systems (e.g., AIRs: autonomous inspection robots) must sustain justified confidence in their correctness and safety—not only at design-time, but also during deployment and throughout post-deployment evolution. Traditional assurance approaches often separate development-time (static) and runtime (dynamic) evidence, leading to fragmented assurance arguments that are difficult to maintain as systems evolve. 
	This challenge is particularly significant in the context of \emph{assured autonomy}, where autonomous systems must maintain high levels of \emph{trustworthiness}~\cite{Abeywickrama-CACM2024} while operating in dynamic and uncertain environments~\cite{Rouff2022}. As autonomy increases, so does the need for assurance arguments capable of \emph{co-evolving} alongside the autonomous system. 
	
	Recent work underscores significant practical challenges in maintaining effective safety cases throughout the lifecycle of autonomous systems. Cârlan et al.~\cite{Carlan2021} identify key shortcomings in current safety case practices: maintenance processes are largely manual, costly, and error-prone; change impact analyses tend to be overly pessimistic; quantitative assessment capabilities remain limited; and only narrow subsets of relevant change scenarios are typically supported. 
	Additionally, automation and traceability between assurance arguments and verification evidence are only partially supported, resulting in limited guidance~\cite{Carlan2021}.
	
	In our study, we distinguish between \emph{static assurance} and \emph{dynamic assurance}~\cite{Schleiss2022,Hartsell2021b}. Static assurance refers to evidence and argumentation generated during development to justify a system’s correctness prior to deployment. By contrast, dynamic assurance concerns operational confidence—i.e., whether the system continues to satisfy its safety requirements at runtime under changing internal or environmental conditions~\cite{Hartsell2021b}. Additionally, we consider \emph{evolution-time assurance} (during post-deployment), which addresses the preservation or restoration of assurance following system modifications (e.g., updates, feature extensions, or reconfiguration)~\cite{Murphy2025}. In our study, we use the term \emph{continuous assurance} to describe this holistic perspective: one that spans design-time, runtime, and evolution. The focus of this paper is specifically on instantiating the design-time stage of the assurance process.
	
	\emph{Formal verification} (e.g., model checking) is the use of mathematical methods to prove or disprove the correctness of a system with respect to a formal specification~\cite{Hoare1969}. 
	In the design-time phase, we employ a heterogeneous verification approach combining two complementary model checking techniques: \emph{RoboChart} with the FDR4 model checker to establish functional correctness and invariant preservation using tock-CSP semantics~\cite{Miyazawa2019}; and \emph{PRISM} to perform probabilistic model checking over Discrete-Time Markov Chains~\cite{PRISM}. By combining RoboChart/FDR4 formal analyses with PRISM’s PCTL (Probabilistic Computational Tree Logic) and reward-based analyses, we obtain a rigorous foundation that supports both logical guarantees and quantitative assessment of key properties of an autonomous system. 
	Meanwhile, \emph{Model-Driven Engineering} (MDE) has emerged as an effective approach for managing assurance artefacts and maintaining traceability as autonomous systems evolve~\cite{Schleiss2022,Belle2023}. 
	Building on MDE principles, we develop a \emph{model-driven assurance pipeline}, implemented as an Eclipse plugin, that automatically transforms formal PRISM artefacts (properties and verification results) into assurance arguments in Goal Structuring Notation (GSN)~\cite{GSN}. Whenever specifications or their verification results change, the plugin automatically regenerates the corresponding assurance arguments, thereby preserving traceability.
	
	This paper makes three main contributions:
	\begin{itemize}
		\item We propose a conceptual framework for \emph{continuous assurance} with three integrated phases—design-time, runtime, and evolution-time—that define artefacts, traceability mechanisms, and automation that contribute to assured autonomy. 
		\item The design-time phase of the framework is instantiated through two \emph{formal verification} workflows—RoboChart/FDR4 for functional analysis, and PRISM for probabilistic analysis. 
		We also propose an MDE approach and plugin that transforms PRISM \emph{specifications} and \emph{verification} results into structured GSN assurance arguments, with automated regeneration when these artefacts change.
		\item We explore our approach using an illustrative nuclear inspection robot scenario. We also discuss it in relation to the joint regulator-issued Trilateral AI Principles~\cite{ONR-TPR_2024}, thereby aligning with regulator-driven best practices for modularity, risk management, secure design, and continuous oversight. 
	\end{itemize}
	This paper is organised as follows. Section~\ref{sec:relatedwork} provides key related work relevant to our study. In Section~\ref{sec:framework}, we describe our proposed framework for continuous assurance. Section~\ref{sec:casestudy} presents the case study involving an illustrative nuclear radiation inspection scenario. Section~\ref{sec:verification} discusses the formal verification performed using RoboChart/FDR4 and PRISM, and outlines the model‑driven pipeline for generating GSN evidence. Section~\ref{sec:discussion} discusses our approach, and Section~\ref{sec:conclusion} concludes with directions for future work.
	
	\section{Related Work}
	\label{sec:relatedwork}
	Continuous assurance for autonomous systems has been explored under several related paradigms, such as \emph{dynamic assurance cases}, \emph{runtime certification}, \emph{adaptive or evolving safety cases}, \emph{through-life assurance}, and more recently \emph{Assurance 2.0 approaches}~\cite{Asaadi2020,BloomfieldRushby2020,Dong2023,Bagheri2023,Avila2023,Belle2023,Schleiss2022,Carlan2021,Jahan2019, Calinescu2018,Denney2015ICSE}. 
	In what follows, we provide key related work around the three lifecycle stages that underpin our framework—design-time, runtime, and evolution-time assurance.
	
	At the \emph{design-time} stage, Bourbouh et al.~\cite{Bourbouh2021} integrate formal verification results into safety arguments via the AdvoCATE tool. However, their method primarily addresses static, design-time evidence and lacks mechanisms for runtime adaptation or automated evolution-time updates. Automation and traceability have also been explored by Wei et al.~\cite{Wei2024}, who embed validation rules directly into GSN models using Constrained Natural Language, enabling executable traceability and consistency checks between assurance arguments and engineering artefacts.
	
	For \emph{runtime} assurance, Belle et al.~\cite{Belle2023} propose dynamic assurance cases for real-time safety assurance in autonomous driving, addressing aleatory uncertainty at design-time and managing epistemic uncertainty during operation through machine learning. Schleiss et al.~\cite{Schleiss2022} present a continuous safety assurance framework that uses runtime monitors to dynamically detect deviations from design-time assumptions, thereby maintaining assurance throughout system operation. 
	Dong et al.~\cite{Dong2023} investigate runtime safety case adaptation, using automated argument modification in response to detected hazards. These approaches reflect growing awareness that static safety arguments risk obsolescence under evolving operational conditions of autonomous systems, but they generally address runtime concerns without strong integration to design-time verification outputs. 
	Several methods have explored dynamically quantifying and updating confidence in safety arguments, including \emph{Bayesian belief networks}~\cite{Denney2011}, \emph{Dempster–Shafer evidence theory}~\cite{WANG2019}, and \emph{Baconian confidence models}~\cite{Weinstock2013}. These frameworks allow assurance arguments to be updated incrementally as new evidence arrives, strengthening or weakening claims based on observed system behaviour. 
	Ferrando et al.~\cite{Ferrando2021} complement these by formalising environmental assumptions as trace expressions, serving both as properties for model checking and as runtime monitors. This trace-based linkage allows violations of design-time assumptions to be detected dynamically. However, these methods do not directly address automation of argument regeneration or integration across lifecycle phases.
	
	At the \emph{evolution-time} phase (post-deployment), methodologies such as ReASSURE~\cite{Murphy2025} support structured change management by identifying which parts of the assurance case are impacted by updates or persistent runtime alerts, prioritising evidence regeneration using pre-assigned cost tags and compositional analysis. This aligns with broader calls for incremental certification and modular safety case maintenance, as emphasised in the survey by Cârlan and Gallina~\cite{Carlan2021}, who highlight the need for modular arguments, precise impact analysis, and enhanced tool support. For self-adaptive and learning-enabled systems, Calinescu et al.~\cite{Calinescu2018} present the ENTRUST methodology, combining design-time and runtime verification for evolving assurance cases. It targets systems whose configurations may change autonomously, thus extending the lifecycle scope of traditional safety arguments.
	
	Our proposed work aims to unify these phases by integrating design-time formal verification, runtime monitoring, and structured evolution-time updates into a holistic process of continuous assurance. Unlike prior work that addresses largely isolated lifecycle phases, we propose a fully interlinked process that guides monitoring, dynamic updates, and regeneration. In doing so, we introduce a continuous assurance framework that enables assurance cases to co-evolve with the autonomous system.
	
	\section{Continuous Assurance Framework}
	\label{sec:framework}
	\begin{figure*}[!t]
		\includegraphics[width=\textwidth]{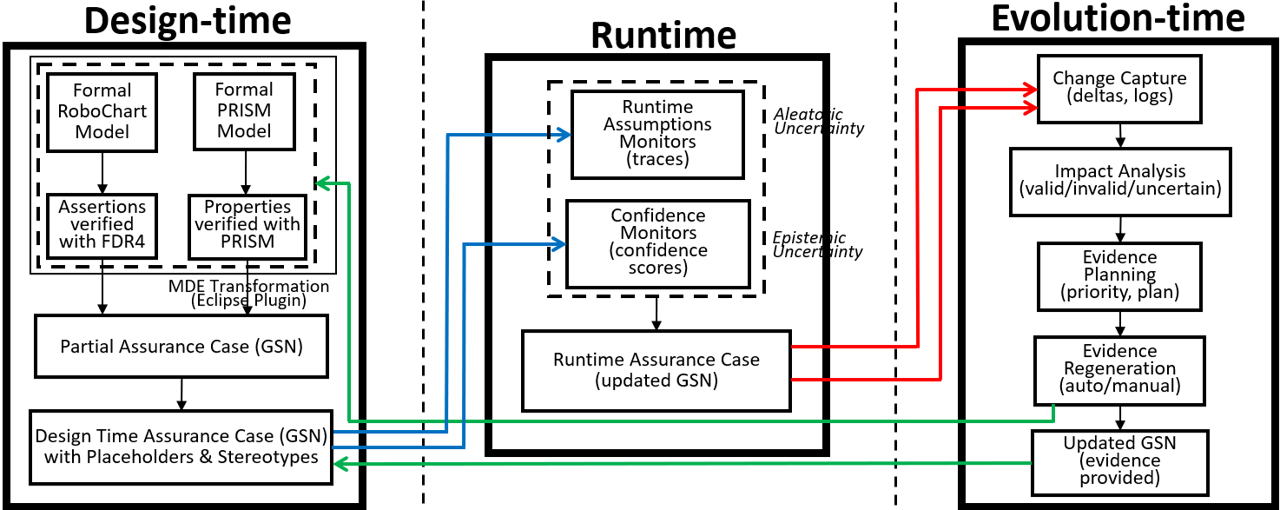}
		\caption{Continuous assurance framework with three stages: design-time, runtime, and evolution-time.} \label{fig:framework}
	\end{figure*}
	This section presents our proposed unified \emph{Continuous Assurance Framework}, which comprises three interlinked phases: \textit{design-time}, \textit{runtime}, and \textit{evolution-time} (see Fig.~\ref{fig:framework}). 
	
	\paragraph*{Design-time assurance (baseline and trace model)} 
	As described earlier, during the design-time phase, the system's behaviour and constraints are formally modelled and verified using two formal verification methods: RoboChart (with the FDR4 model checker) and PRISM. 
	Based on these formal analyses, a \emph{partial assurance case} is created using GSN (Fig.~\ref{fig:framework}). This serves as the baseline for continuous assurance and is automatically updated when the PRISM specification (properties) and verification results (evidence) change, while updates originating from RoboChart/FDR4 analysis are linked manually (see Section~\ref{sec:transformation}).
	
	To support traceability and automated updates, we embed two types of structured annotations directly into GSN goal descriptions: \emph{placeholders} and \emph{stereotypes} (an overview of all placeholders and stereotypes is provided in Table~\ref{tab:placeholders_stereotypes}).  
	Placeholders encode deferred claims that must be validated at runtime or post-deployment (e.g., \texttt{trace\_expr="..."}, \texttt{monitor\_expr="..."}). For example, \texttt{trace\_expr} captures environmental or control assumptions that guide runtime monitor configuration and violation detection. Stereotypes (e.g., \texttt{<<TraceMonitored>>}, \texttt{<<DeferredEvidence>>}) provide a lightweight classification mechanism to classify nodes, support filtering and enable lifecycle-aware modelling. 
	
	The introduction of \emph{placeholders} and \emph{stereotypes} into the GSN model provides explicit, machine-interpretable link points that can be exploited by analysis plugins and runtime monitoring components. These annotations allow automated tools to identify which claims rely on deferred evidence, which assumptions are actively monitored, and which argument nodes are affected by changes in verification results or operational data. 
	By embedding this metadata, the framework enables targeted evidence retrieval, selective re-verification, and focused runtime monitoring—avoiding exhaustive manual re-analysis. 
	In practice, placeholders become effective trace links once populated with concrete identifiers (e.g., PRISM model artefacts), enabling systematic forward propagation of updates into the corresponding GSN nodes and providing a foundation for structured change management. This trace infrastructure supports variability-aware workflows~\cite{Murphy2025} and ensures that the assurance argument remains tightly coupled to the verified system model—even as system requirements or configurations evolve.
	
	\paragraph*{Runtime assurance (dynamic monitoring and evidence update)}  
	
	During operation, the system may encounter both \emph{aleatory} and \emph{epistemic} uncertainty~\cite{Schleiss2022} that can invalidate assumptions established at design-time. To maintain assurance under these conditions, we incorporate two types of runtime monitors into the GSN model—\emph{runtime assumption monitors} and \emph{confidence monitors}—each defined using structured placeholders and stereotypes. Runtime assumption monitors address violations typically stemming from aleatory uncertainty (random noise, data gaps), whereas confidence monitors help manage epistemic uncertainty (previously unknown risks revealed during operation)~\cite{Schleiss2022}.
	
	\emph{Runtime assumption monitors} detect violations of assumptions about sensor data or control-loop behaviour established during design. Each monitor is represented by annotating a GSN goal node with the \texttt{<<RuntimeAssumptionMonitor>>} stereotype and a description placeholder (e.g., \texttt{monitor\_expr="..."}). Each monitor is also assigned a \texttt{monitor\_id} to link GSN nodes with runtime logs and mitigation actions. When a violation occurs, the corresponding node is marked with \texttt{<<Reopened>>} and populated with a runtime log or mitigation response.
	
	In parallel, \emph{confidence monitors} track a numerical ``confidence score'' for each GSN goal, inspired by dynamic assurance case approaches that embed probabilistic reasoning within structured arguments. Prior work has explored mapping safety/assurance arguments to Bayesian belief networks for evidence updating~\cite{Denney2011}, applying Dempster--Shafer belief functions to combine heterogeneous evidence~\cite{WANG2019}, and using Baconian-style eliminative reasoning to quantify confidence~\cite{Weinstock2013}. 
	
	\begin{sloppypar}
		In our model, each confidence monitor is represented by annotating a GSN goal node with the \texttt{<<ConfidenceMonitor>>} stereotype and a description placeholder (e.g., \texttt{confidence\_threshold="0.95"}) (Table~\ref{tab:placeholders_stereotypes}). If confidence drops below this threshold, the goal is reopened and marked with \texttt{<<DeferredEvidence>>} for review. All monitor outputs (e.g., events, logs, confidence scores) are trace-linked to corresponding GSN nodes, maintaining consistency between runtime evidence and the assurance argument.
		This structured use of placeholders and stereotypes ensures the dynamic assurance case remains continuously updated—reinforcing, weakening, or reopening claims in response to operational conditions.
	\end{sloppypar}
	
	\paragraph*{Evolution-time assurance (change management and regeneration)}  
	The evolution-time phase adopts a structured change management process inspired by ReASSURE~\cite{Murphy2025}. When system changes arise—such as software updates, reconfigurations, feature additions, or persistent runtime alerts—GSN nodes previously marked with \texttt{<<DeferredEvidence>>} or \texttt{<<ConfidenceMonitor>>} serve as entry points for analysis (Table~\ref{tab:placeholders_stereotypes}). This can be performed by assembling an \emph{evolution package} comprising model deltas, monitor logs, incident reports, and reopened GSN goals. Each opened goal carries an \texttt{evidence\_cost="..."} tag from design-time to estimate the required regeneration effort. 
	Using the \texttt{evidence\_cost} tags and trace links from each GSN node to its originating RoboChart or PRISM artefact, we generate an \texttt{<<ImpactAnalysis>>} summary classifying goals as \emph{valid}, \emph{invalid}, or \emph{uncertain}. Invalid or uncertain goals become candidates for regeneration. During the evidence planning phase, we prioritise these based on \texttt{evidence\_cost} and safety criticality, and annotate each with a \texttt{<<RegenerationPlan>>} stereotype describing whether proof, simulation, or manual review is required.
	
	Evidence regeneration can be performed using automated tools (e.g., rerunning CSP/FDR4 checks or PRISM analyses) for low-cost goals, while complex or high-impact goals can be addressed through manual modelling and review. Upon successful regeneration, each affected GSN node’s placeholder is updated with new evidence, its stereotype is changed from \texttt{<<DeferredEvidence>>} to \texttt{<<EvidenceProvided>>}, and its version tag is incremented. Runtime monitor thresholds (from \texttt{<<RuntimeAssumptionMonitor>>} stereotypes) are also restored or adjusted based on the new evidence.
	\begin{table*}[!t]
		\centering
		\caption{Use of placeholders and stereotypes across design-time, runtime, and evolution-time assurance.}
		\label{tab:placeholders_stereotypes}
		\renewcommand{\arraystretch}{1.3}\small
		\begin{tabular}{|p{4.8cm}|p{1.8cm}|p{1.7cm}|p{6.5cm}|}
			\hline
			\textbf{Name} & \textbf{Phase(s)} & \textbf{Type} & \textbf{Description} \\
			\hline
			\texttt{trace_expr="..."} & Design, \mbox{Runtime} & Placeholder & Encodes trace-based assumptions for monitor configuration and violation detection. \\
			\hline
			\texttt{monitor_id="..."} & All phases & Placeholder & Unique identifier linking GSN goal to runtime monitor and logs. \\
			\hline
			\texttt{deferred=true} & All phases & Placeholder & Indicates that claim is not discharged yet and requires runtime or post-change validation. \\
			\hline
			\texttt{confidence_threshold="..."} & Runtime & Placeholder & Minimum accepted confidence score before a goal is reopened. \\
			\hline
			\texttt{monitor_expr="..."} & Runtime & Placeholder & Specifies runtime condition or logic checked by a monitor. \\
			\hline
			\texttt{evidence_cost="..."} & Design, \mbox{Evolution} & Placeholder & Time/effort estimate to regenerate evidence. Used in prioritisation. \\
			\hline
			\texttt{evolution_package="..."} & Evolution & Placeholder & Contains triggering artifacts: logs, diffs, reports. \\
			\hline
			\texttt{impact_summary="..."} & Evolution & Placeholder & Classifies analysis results: valid / invalid / uncertain. \\
			\hline
			\texttt{regeneration_plan="..."} & Evolution & Placeholder & Specifies repair strategy: simulation, proof, manual. \\
			\hline
			\texttt{<<TraceMonitored>>} & Design, \mbox{Runtime} & Stereotype & Identifies goals linked to runtime-monitored trace assumptions. \\
			\hline
			\texttt{<<DeferredEvidence>>} & All phases & Stereotype & Flags goals where evidence is postponed, missing, or reopened. \\
			\hline
			\texttt{<<RuntimeAssumption\allowbreak Monitor>>} & Runtime, \mbox{Evolution} & Stereotype & Identifies runtime assumption monitors linked to control or sensor assumptions. \\
			\hline
			\texttt{<<ConfidenceMonitor>>} & Runtime, \mbox{Evolution} & Stereotype & Tracks probabilistic/confidence-based monitors. \\
			\hline
			\texttt{<<Reopened>>} & Runtime, \mbox{Evolution} & Stereotype & Marks goals reopened due to threshold violation or runtime alert. \\
			\hline
			\texttt{<<RegenerationPlan>>} & Evolution & Stereotype & Indicates plan to recover missing evidence. \\
			\hline
			\texttt{<<ImpactAnalysis>>} & Evolution & Stereotype & Tags nodes summarizing outcome of change analysis. \\
			\hline
			\texttt{<<EvidenceProvided>>} & Evolution & Stereotype & Confirms goal closure with new evidence after regeneration. \\
			\hline
		\end{tabular}
	\end{table*}
	\normalsize
	
	This essentially closes the feedback loop, ensuring that evolution-time activities resolve the placeholders and stereotypes introduced at design-time, and maintain alignment between the assurance case and the autonomous system as it evolves. 
	Our framework supports partially automated assurance case maintenance across the lifecycle, where automation can be used to support low-impact claims, while manual intervention can be reserved for high-stakes scenarios. 
	By unifying specifications, verification results, and assurance case regeneration in a single workflow, our framework aims to maintain a consistent \emph{single source of truth} from the formal model to structured argument, thereby contributing to agile and systematic safety case maintenance.
	
	\section{Case Study: An Illustrative Scenario of a Nuclear Radiation Inspection Robot}
	\label{sec:casestudy}
	We illustrate our continuous assurance approach through a case study involving a \emph{Nuclear Radiation Inspection Robot}, focusing on the following mission scenario. An unmanned ground vehicle (UGV), such as a Scout Mini, is deployed to patrol a sequence of four predefined waypoints (e.g., \(l_{3}\), \(l_{4}\), \(l_{1}\), and \(l_{2}\)) within a nuclear storage facility. 
	At each stop, the UGV must measure ambient radiation levels, timestamp its dose-rate readings, and transmit the data to a remote monitoring station.  
	The UGV is equipped with LiDAR for obstacle detection, depth cameras for localization, radiation sensors for dose-rate measurement, and onboard diagnostics for battery monitoring.  
	
	To ensure mission integrity and safety, the UGV must: (a) follow a strict, ordered patrol route without deviation (e.g., \(l_{3} \rightarrow l_{4} \rightarrow l_{1} \rightarrow l_{2}\)); (b) avoid restricted or ``no-go'' zones (e.g., known radiation hot spots); and (c) terminate the mission early if battery charge falls below a safe operational threshold.  
	To enforce these requirements under elevated radiation, a three-state \textit{safety-wrapper controller} supervises the UGV's operational mode—\texttt{Patrol}, \texttt{Caution}, and \texttt{Emergency Retrieval}. Each ambient dose-rate reading is evaluated against two thresholds, \(R_{1}\) and \(R_{2}\): if the dose exceeds \(R_{1}\), the controller switches to \texttt{Caution} mode and slows the UGV; if it exceeds \(R_{2}\), it initiates an immediate transition to \texttt{Emergency Retrieval}. In \texttt{Caution} mode, the robot continues along its waypoint plan at reduced speed to limit radiation exposure and conserve energy. In \texttt{Emergency Retrieval} mode, the robot halts in place (absorbing behaviour) and awaits recovery or manual retrieval. Battery depletion below the defined threshold in any mode triggers an \texttt{Abort} state, ending the mission safely. In both \texttt{Caution} and \texttt{Emergency Retrieval} modes, operator or navigator inputs are ignored to prevent unsafe overrides.  
	This scenario serves as the common reference model for both our RoboChart and PRISM verification workflows, supporting formal analysis of functional correctness, safety guarantees, and probabilistic mission outcomes.
	
	\section{Formal Verification using RoboChart and PRISM}
	\label{sec:verification}
	In this section, we describe the design-time phase of our framework. Specifically, we present two \emph{formal verification} workflows—RoboChart/FDR4 for functional analysis, and PRISM for probabilistic analysis—and show how assurance artefacts are regenerated when requirements or verification results change.
	
	Formal verification is a central element of our assurance methodology, integrated into the development process under a \emph{design-for-assurance} paradigm, providing mathematical evidence that critical safety properties are satisfied by design before deployment.
	Unlike simulation or testing—which explore only a finite subset of behaviours—formal verification techniques such as model checking examine all reachable executions as determined by a model’s formal semantics. This is particularly important for autonomous systems, where uncommon environmental conditions and rare concurrency interleavings can trigger failures that conventional testing misses. 
	Our verification approach employs a \emph{heterogeneous verification} strategy~\cite{Luckcuck2019} using two complementary toolchains: RoboChart with FDR4 for refinement-based functional correctness, and PRISM for analysis under probabilistic uncertainty. Together, these enable both verification of safety properties over all executions and quantitative assessments of the likelihood of undesirable events under realistic operational assumptions.

	\subsection{RoboChart Modelling of the Case Study}
	We first model the core safety behaviour of the inspection robot’s control architecture using RoboChart~\cite{Miyazawa2019,Wei2024b}, a domain-specific modelling language based on state machines with semantics defined in CSP. RoboChart models are organised into modules, each containing a robotic platform and one or more concurrent controllers. Using RoboTool, RoboChart specifications are translated into CSP under tock-CSP semantics~\cite{Miyazawa2019}, enabling timed refinement checking in the FDR4 model checker. 
	\begin{figure*}[!t]\centering
		\includegraphics[width=0.9\textwidth]{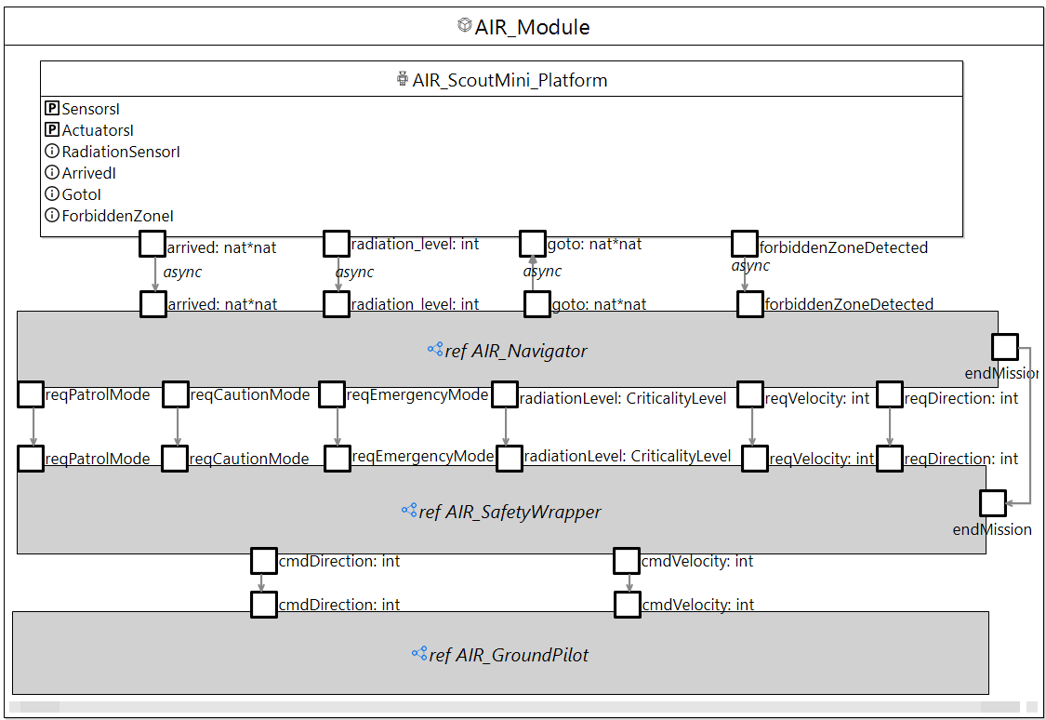}
		\caption{High-level architecture of the nuclear inspection robot and safety wrapper in the case study scenario.} \label{fig:RC1}
	\end{figure*}
	
	In the case study (Section~\ref{sec:casestudy}), we model the high-level safety architecture of an unmanned ground vehicle (UGV) for nuclear radiation inspection. Figure~\ref{fig:RC1} shows the architecture comprising three controllers: \texttt{AIR_Navigator} (responsible for issuing high-level motion commands and detecting radiation levels), \texttt{AIR_SafetyWrapper} (responsible for enforcing radiation-aware safety policies), and \texttt{AIR_GroundPilot} (responsible for executing approved commands on the robot’s actuators). \texttt{AIR_ScoutMini_Platform}, which is the robotic platform, acts as a hardware abstraction layer providing access to sensors and actuators. The \texttt{AIR_SafetyWrapper} acts as a mediating safety layer: it intercepts operator (or navigator) commands, evaluates them against the current hazard mode, and either forwards or suppresses them. This centralisation of safety logic enforces a clear separation of concerns and facilitates traceability from requirements to model elements.
	
	\begin{figure*}[!t]\centering
		\includegraphics[width=0.9\textwidth]{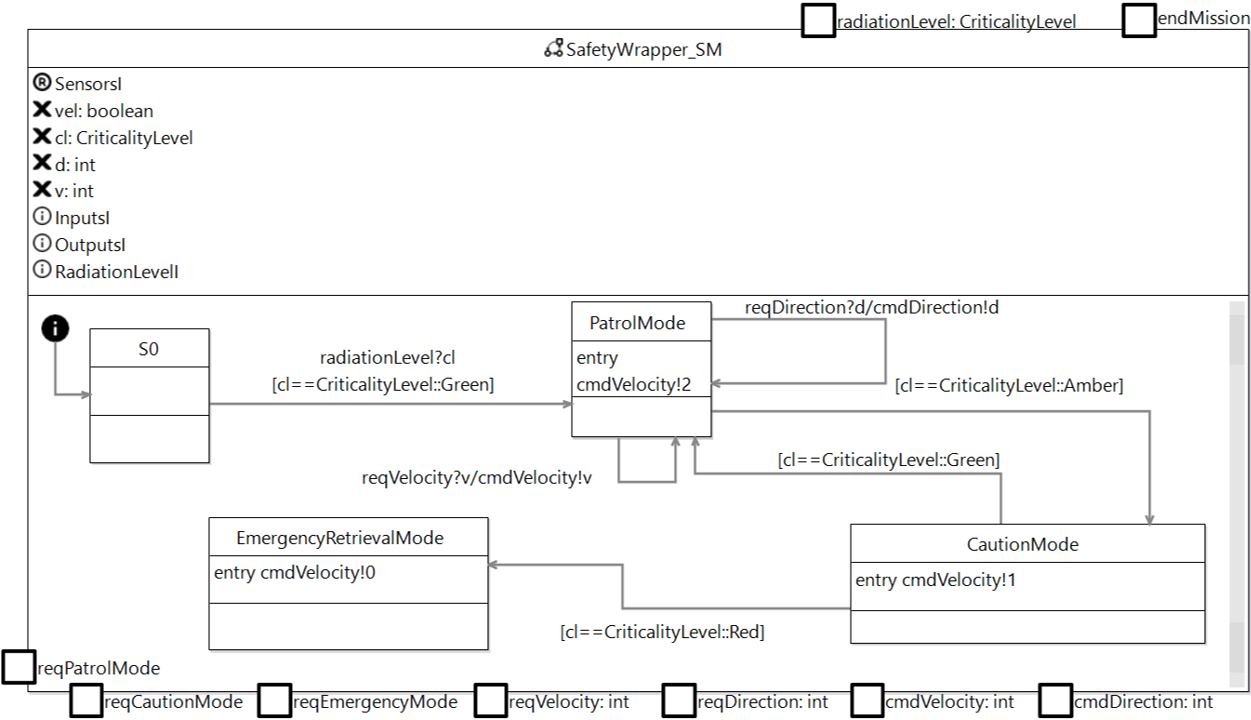}
		\caption{State machine model of the safety wrapper controller in the case study scenario.} \label{fig:RC2}    
	\end{figure*}
	
	\begin{figure*}[!t]\centering
		\includegraphics[width=0.9\textwidth]{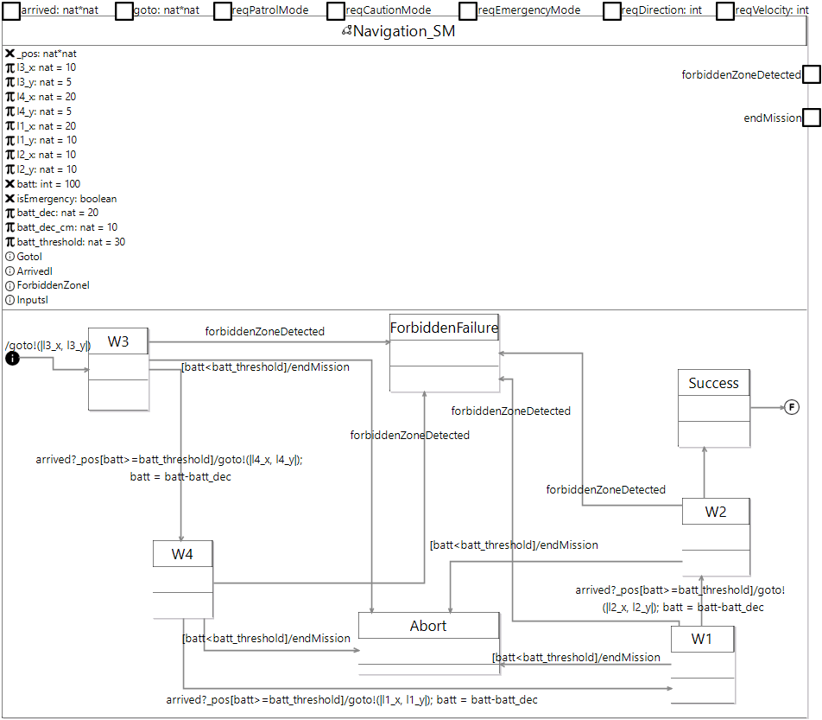}
		\caption{Navigation state machine model for waypoint navigation in the case study scenario.} \label{fig:RC3}    
	\end{figure*}
	At the core of the \texttt{AIR_SafetyWrapper} is a hierarchical three-mode state machine (see Fig.~\ref{fig:RC2}) that adapts to ambient radiation levels. In \emph{Patrol Mode (PM)}, the robot follows a waypoint sequence at full speed while blocking entry into unsafe zones. When the dose exceeds threshold $R_1$, the wrapper transitions to \emph{Caution Mode (CM)}, reducing speed and raising alerts. If the dose exceeds the critical threshold $R_2$, the system switches to \emph{Emergency Retrieval Mode (ERM)}, an absorbing safety state where the robot halts in place and awaits recovery or manual retrieval. The speed adjustments are executed by the \texttt{AIR_GroundPilot} controller via the robot’s actuators. 
	
	Meanwhile, the \texttt{AIR_Navigator} controller comprises two state machines: \texttt{Navigation_SM} (see Fig.~\ref{fig:RC3}), and \texttt{RadiationLevelMonitor_SM}. 
	The \texttt{Navigation_SM} defines the waypoint-based mission behaviour, capturing both its progression through the patrol route and the safety constraints that can terminate the mission early. This state machine specifies four sequential waypoint states (\texttt{W3}, \texttt{W4}, \texttt{W1}, \texttt{W2}) connected by transitions guarded by arrival events and battery thresholds. At each transition, the robot issues a \texttt{goto} command to the next waypoint and decrements the battery level accordingly. In addition to this nominal route, \texttt{Navigation_SM} introduces safety-related terminal states: \texttt{Success} (mission completion after reaching the final waypoint), \texttt{ForbiddenFailure} (triggered by entry into restricted zones), and \texttt{Abort} (triggered by insufficient battery charge). 
	The \texttt{RadiationLevelMonitor_SM} models the robot’s handling of radiation sensor inputs and their translation into abstract hazard levels for the safety wrapper (i.e., \texttt{Green} (safe), \texttt{Amber} (warning), and \texttt{Red} (critical)). 
	The \texttt{AIR_Navigator}, \texttt{AIR_SafetyWrapper}, and \texttt{AIR_GroundPilot} processes are then composed in parallel, synchronising on shared channels to form a closed-loop control model.
	
\begin{figure*}[!t]
		\includegraphics[width=\textwidth]{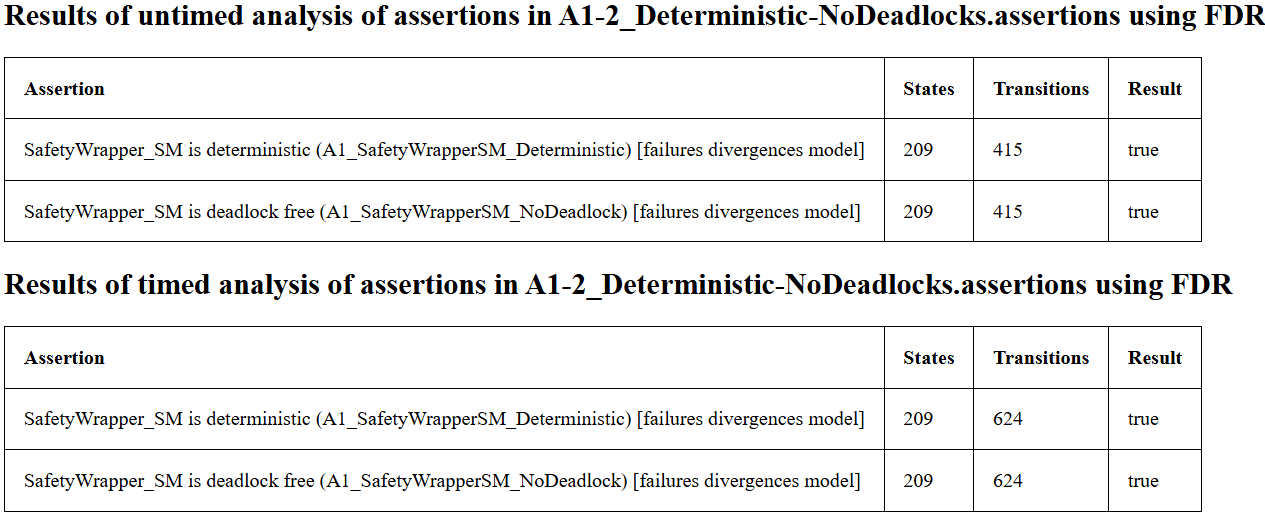}
		\caption{Verifying that the SafetyWrapper_SM state machine is deadlock-free and deterministic.} \label{fig:rc-V}
	\end{figure*}

	We formalise the system requirements as CSP refinement assertions over observable controller events and incorporate them into the CSP specifications generated by RoboTool. These assertions capture high-level requirements covering both safety and liveness properties. For example, they ensure that the system maintains progress, follows its strict patrol route, avoids forbidden zones, responds appropriately to radiation thresholds, blocks unsafe inputs, and adheres to battery and emergency constraints. Any violation produces a counterexample trace that identifies the specific sequence of events leading to failure. Below, we present several key assertions for the \texttt{SafetyWrapper_SM} state machine and illustrate one in Listing~\ref{lst:modeswitch}. Figure~\ref{fig:rc-V} shows the result of verifying that \texttt{SafetyWrapper_SM} is deadlock-free and deterministic.
	
	\begin{itemize}
		\item \textbf{A1-2\_Deterministic-NoDeadlocks} – The system (e.g., \texttt{SafetyWrapper\_SM}) must be deterministic. Also, \texttt{SafetyWrapper\_SM}) must never enter a state with no further progress possible.
		\item \textbf{A3\_WaypointSeq} – Ensures that the UGV visits waypoints in strict sequence.
		\item \textbf{A4\_ModeSwitch} – Ensures correct mode transitions in response to radiation readings.
		\item \textbf{A5\_Velocity} – Enforces immediate velocity adjustment on hazard detection.
		\item \textbf{A6\_NoReq} – Blocks operator and navigator control inputs during unsafe modes (CM or ERM).
		\item \textbf{A7\_ERMAbsorb} – Once in Emergency Retrieval Mode, the system must remain there until mission termination.
		\item \textbf{A8\_BattAbort} – If battery charge falls below the configured threshold, the system must transition immediately to Abort and end the mission.
	\end{itemize}
	
	\paragraph*{Mode–switch correctness under radiation readings}
	\texttt{A4_ModeSwitch}, a CSP refinement assertion, ensures that each radiation reading is followed by the appropriate speed command by enforcing the mapping between detected radiation levels and the corresponding commanded velocities (see Listing~\ref{lst:modeswitch}).
\begin{lstlisting}[caption={Mode–switch specification and refinement assertion.},label = lst:modeswitch]
csp ModeSwitchSpec associated to SafetyWrapper_SM
csp-begin
ModeSwitchSpec =
SafetyWrapper_SM::radiationLevel.in?cl ->
( if cl == CriticalityLevel_Green
then SafetyWrapper_SM::cmdVelocity.out!2 -> ModeSwitchSpec
else if cl == CriticalityLevel_Amber
then SafetyWrapper_SM::cmdVelocity.out!1 -> ModeSwitchSpec
else
SafetyWrapper_SM::cmdVelocity.out!0 -> ModeSwitchSpec )
|~| SafetyWrapper_SM::reqDirection.in?d   -> ModeSwitchSpec
|~| SafetyWrapper_SM::reqPatrolMode.in    -> ModeSwitchSpec
|~| SafetyWrapper_SM::endMission.in       -> ModeSwitchSpec
|~| SafetyWrapper_SM::reqCautionMode.in   -> ModeSwitchSpec
|~| SafetyWrapper_SM::reqVelocity.in?v    -> ModeSwitchSpec
|~| SafetyWrapper_SM::reqEmergencyMode.in -> ModeSwitchSpec
|~| SafetyWrapper_SM::cmdDirection.out?d  -> ModeSwitchSpec
|~| SafetyWrapper_SM::terminate           -> SKIP
csp-end
untimed assertion A4_ModeSwitch: SafetyWrapper_SM refines ModeSwitchSpec in the traces model
\end{lstlisting}
	
	\subsection{PRISM Modelling and Verification of the Case Study}
	
	While RoboChart and FDR4 provide exhaustive guarantees of logical and real-time correctness, they do not account for stochastic disturbances, including random collisions, radiation spikes, and energy depletion. To address these aspects, we construct a formal model in PRISM~\cite{PRISM}, as a Discrete-Time Markov Chain (DTMC), parameterised by environmental factors such as the probability of entering a forbidden zone ($p_{\mathit{err}}$) and radiation event rates. This enables both qualitative and quantitative verification under probabilistic behaviour. Our current PRISM model for the case study comprises two synchronised modules: \texttt{AIR\_Navigator}, which captures the mission logic including waypoint navigation, battery usage, and radiation sampling; and \texttt{AIR\_SafetyWrapper}, which enforces runtime safety policies through a three-mode supervisory controller. The wrapper regulates both the robot’s velocity and operational mode based on sensed radiation levels. The PRISM model of the case study scenario is provided in Appendix A.
	
	\begin{table*}[!t]
		\centering
		\footnotesize
		\caption{PCTL and reward-based properties defined for the PRISM model of the case study.}
		\label{tab:prism-properties}
		\begin{tabular}{|l|p{5cm}|p{7.5cm}|}
			\hline
			\textbf{Name} & \textbf{Property (PCTL/Reward)} & \textbf{Description} \\
			\hline
			\multicolumn{3}{|l|}{\textbf{Navigation Properties}} \\
			\hline
			P\_succ        & $P=?\ [F\ \text{loc} = 4]$  & Probability that the mission eventually succeeds \\
			P\_forb        & $P=?\ [F\ \text{loc} = 5]$  & Probability that the robot enters a forbidden zone eventually \\
			P\_safe        & $P=?\ [G\ \text{loc} != 5]$ & Probability that the mission always avoids a forbidden zone \\
			P\_condSucc    & $P=?\ [(\text{loc} != 5)\ U\ (\text{loc} = 4)]$ & Probability that the goal is reached without entering a forbidden zone. \\
			P\_battRisk    & $P=?\ [F\ \text{batt} < \text{batt\_threshold}]$ & Probability that battery eventually drops below the minimum safety threshold.\\
			P\_timeBound   & $P=?\ [F\ <= 5\ \text{loc} = 4]$ & Probability that the mission completes within 5 navigation steps \\
			P\_safeEnergy  & $P=?\ [G\ (\text{loc} != 5\ \&\ \text{loc} != 6\ \&\ \text{batt} >= \text{batt\_threshold})]$ & Probability that the robot always remains safe and above energy threshold. \\
			R\_dose        & $R\{"\text{dose}"\}=?\ [F\ (\text{loc} = 4\ |\ \text{loc} = 5\ |\ \text{loc} = 6)]$ & Expected cumulative number of radiation exposure events before mission termination \\
			R\_moves       & $R\{"\text{moves}"\}=?\ [F\ (\text{loc} = 4\ |\ \text{loc} = 5\ |\ \text{loc} = 6)]$ & Expected total number of navigation moves until absorption (success, failure, or abort). \\
			R\_time\_cm    & $R\{"\text{time\_in\_cm}"\}=?\ [F\ (\text{loc} = 4\ |\ \text{loc} = 5\ |\ \text{loc} = 6)]$ &  Expected cumulative time steps spent in Caution mode before mission termination. \\
			R\_time\_stopped & $R\{"\text{time\_stopped}"\}=?\ [F\ (\text{loc} = 4\ |\ \text{loc} = 5\ |\ \text{loc} = 6)]$ & Expected cumulative time steps halted in Emergency Retrieval mode. \\
			\hline
			\multicolumn{3}{|l|}{\textbf{Safety Wrapper Properties}} \\
			\hline
			P\_warnMode    & $P >= 1\ [F\ (\text{rad} = 1\ \&\ \text{sw} = 1)]$ & Guarantee that warning-level radiation always triggers entry into Caution mode. \\
			P\_critMode    & $P >= 1\ [F\ (\text{rad} = 2\ \&\ \text{sw} = 2)]$ & Guarantee that critical-level radiation always triggers transition into Emergency Retrieval mode \\
			P\_fullSpeed   & $P >= 1\ [G\ (\text{sw} = 0\ ->\ \text{vel} = 2)]$ & Guarantee that Patrol mode always enforces full velocity. \\
			P\_slowSpeed   & $P >= 1\ [G\ (\text{sw} = 1\ ->\ \text{vel} = 1)]$ & Guarantee that Caution mode always enforces reduced velocity \\
			P\_stopped     & $P >= 1\ [G\ (\text{sw} = 2\ ->\ \text{vel} = 0)]$ & Guarantee that Emergency Retrieval mode always enforces a complete stop  \\
			P\_noOpOutside & $P <= 0\ [F\ (\text{sw} != 0\ \&\ \text{op\_used})]$ & Guarantee that operator input never occurs outside Patrol mode \\
			\hline
		\end{tabular}
	\end{table*}
	
	The robot navigates through four sequential waypoints (\texttt{loc} = 0 to 3), with terminal states defined as \texttt{loc} = 4 for successful completion, \texttt{loc} = 5 for forbidden-zone entry, and \texttt{loc} = 6 for battery-abort or completed emergency retrieval. Radiation levels are discretised into three categories: Safe (0), Warning (1), and Critical (2). The battery state-of-charge (\texttt{batt}) ranges from 0 to 100, decreasing with each movement. 
	The control mode variable \texttt{sw} indicates the current safety state: 0 = Patrol, 1 = Caution, and 2 = Emergency retrieval. The \texttt{AIR\_Navigator} module governs transitions through the waypoints, where each step consumes energy, involves a risk of entering a forbidden zone (\texttt{p\_err}), and results in probabilistic radiation sampling. If \texttt{batt} falls below a defined threshold, the robot transitions to the abort state.
	
\begin{figure*}[!t]
		\centering
		\includegraphics[width=0.7\textwidth]{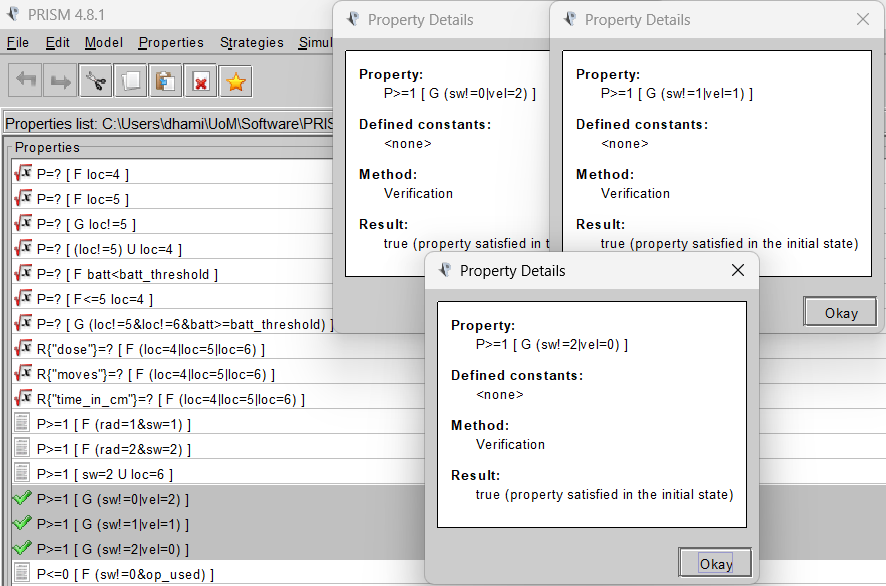}
		\caption{Formal verification of PRISM properties in the case study.}
		\label{fig:prism_r}
	\end{figure*}

	The \texttt{AIR\_SafetyWrapper} manages mode transitions in response to environmental conditions. Warning-level radiation triggers a transition from \texttt{Patrol} to \texttt{Caution}, reducing speed. 
	Critical radiation immediately switches the system to \texttt{Emergency Retrieval} mode, an absorbing state in which the robot halts in place and awaits recovery or manual retrieval. 
	All transitions to Emergency Retrieval are absorbing; once entered, the robot remains in safety override mode. Additionally, operator inputs (via the synchronised labels \texttt{[hdng]} and \texttt{[vel]}) are permitted only in \texttt{Patrol} mode and are tracked using the Boolean variable \texttt{op\_used}. 
	To assess the system quantitatively, we define several reward structures: \texttt{moves} (forward progress), \texttt{dose} (high-radiation exposures), \texttt{time\_in\_cm} (duration in \texttt{Caution} mode), and \texttt{time\_stopped} (halted states in \texttt{Emergency Retrieval}). These allow cumulative tracking of operational behaviours relevant to the mission and safety assurance. A comprehensive set of probabilistic and reward-based properties in PCTL (see Table~\ref{tab:prism-properties} and Fig.~\ref{fig:prism_r}) is used to verify correctness, assess risks, and confirm enforcement of several key safety policies at runtime.
	
	\subsection{Transforming Formal Specification and Verification Results into Assurance Artefacts}
	\label{sec:transformation}
	As autonomous systems evolve, their structured assurance case arguments must remain synchronised with both the underlying formal specifications and associated verification results (design-time evidence). This traceability is central, as manual regeneration of assurance artefacts introduces risks of inconsistency and transcription error, and can hinder agile development. Towards addressing this, we developed \texttt{PRISM2GSN}, an Eclipse plugin based on the Eclipse Rich Client Platform (RCP) and OSGi component model, which integrates the PRISM model checker into the Eclipse development environment and automates the regeneration of GSN arguments from updated property specifications. The plugin is available for download from~\cite{PRISM2GSN-GitHub} along with instructions for its use.
	
	The \texttt{PRISM2GSN} plugin monitors temporal logic property specifications in PRISM \texttt{.props} files, which may include probabilistic reachability properties, time-bounded properties, and reward-based properties. 
	Whenever a property file is saved, a verification-to-assurance pipeline is triggered automatically: the relevant PRISM model is executed, results are parsed, and the corresponding assurance arguments in GSN are regenerated. 
	Internally, a lightweight \texttt{Activator} class registers a resource change listener that responds to updates in \texttt{.props} files. 
	Upon detection, a background \texttt{WorkspaceJob} is scheduled to safely perform the transformation without causing file-lock contention. 
	This job invokes the \texttt{Prism2GSNTransformer}, which locates the associated \texttt{.prism} model in the project directory and executes the PRISM command-line interface via \texttt{cmd.exe} in the PRISM \texttt{bin} directory. 
	The location of this directory is configurable via the Eclipse Preferences UI, allowing the plugin to flexibly support different PRISM installations. 
	Standard output is captured and parsed using regular expressions to extract verification results, including Boolean verdicts, probabilities, and expected rewards. The plugin currently handles a range of PCTL property types, including time-bounded, conditional, and reward-based queries, and is extensible to accommodate new property classes.

\begin{figure*}[!t]
		\includegraphics[width=\textwidth]{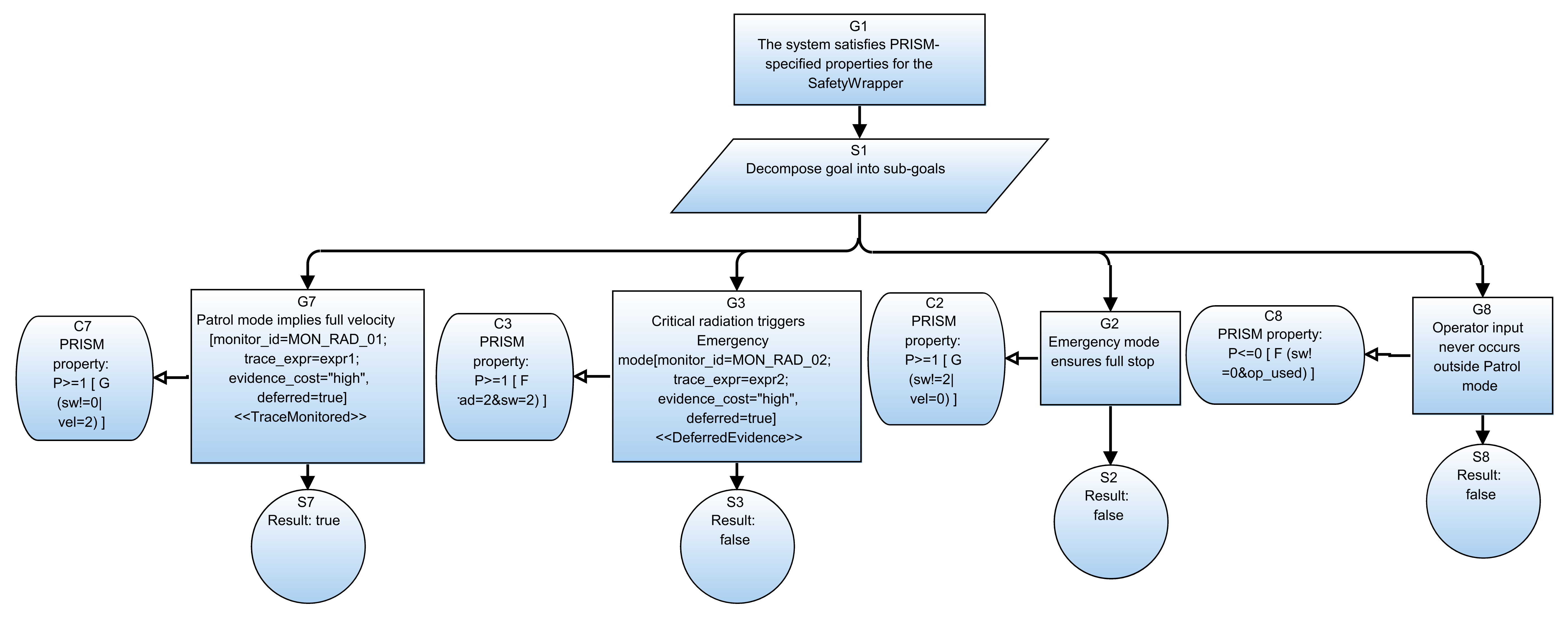}
		\caption{Part of an assurance case generated by the PRISM2GSN plugin for the safety wrapper, with several placeholders and stereotypes manually inserted at design-time after generation.} \label{fig:assurancecase}
	\end{figure*}

	The extracted results are transformed into a structured assurance argument in the domain-specific language (DSL) of the AdvoCATE assurance case tool~\cite{Denney2023,AdvoCATE-UG}. 
	Currently, a generated argument contains (i) a top-level goal asserting satisfaction of the verified property set; a strategy node describing decomposition by individual property; and one sub-goal per property, each linked to its formal specification (context) and PRISM verification result (solution). The generated DSL file can be directly visualised and edited in the AdvoCATE tool. 
	After generating assurance arguments, structured placeholders and stereotypes are manually inserted into the GSN elements to support traceability and enable automated updates during runtime and post-deployment. 
	Figure~\ref{fig:assurancecase} shows part of an assurance case generated by the PRISM2GSN plugin for the safety wrapper, with several placeholders and stereotypes (e.g., \texttt{trace_expr}, \texttt{evidence_cost}, \texttt{<<DeferredEvidence>>}) manually inserted at design-time after generation.
	
	\section{Discussion: Regulatory Considerations}
	\label{sec:discussion}
	This section provides a discussion of our proposed approach, focusing on its alignment with regulatory expectations. The \emph{Trilateral Principles} study~\cite{ONR-TPR_2024}, jointly published by the Canadian Nuclear Safety Commission (CNSC), the United Kingdom’s Office for Nuclear Regulation (ONR), and the United States Nuclear Regulatory Commission (NRC), outlines regulatory expectations for the safe deployment of AI in nuclear applications across five key domains: safety and security engineering, human factors, AI architecture, lifecycle management, and assurance documentation. Our study examines these regulatory expectations, thereby reflecting regulatory alignment.

	\textit{Risk-Proportionate Design and Mode Stratification:}
	The risk model in~\cite[Sec.~3]{ONR-TPR_2024} advocates proportionate assurance aligned with both autonomy level and safety consequence. Our study implements a three-mode \texttt{AIR\_SafetyWrapper}—\texttt{Patrol}, \texttt{Caution}, and \texttt{Emergency Retrieval}—which directly correspond to Regions~2--4 of~\cite[Sec.~3]{ONR-TPR_2024}, representing increasing autonomy and consequence. \texttt{Patrol} mode supports routine low-dose surveying with full autonomy; \texttt{Caution} mode enforces speed limitations under moderate radiation; and \texttt{Emergency Retrieval} triggers immediate withdrawal upon high-dose detection.
	
	\textit{Modular and Secure AI Architecture:}  
	In alignment with~\cite[Secs.~6, 4]{ONR-TPR_2024}, we adopt a modular design with clearly defined component boundaries and constrained input/output interfaces. All radiation-response logic is encapsulated within a standalone \texttt{AIR\_SafetyWrapper} module that intercepts and regulates commands issued by the operator or the navigator. The module is formally verified to ensure deadlock freedom, compliance with mode-transition deadlines, and enforcement of operational constraints. The architecture preserves separation of concerns by decoupling AI-based decision logic from actuation and communication subsystems, consistent with the report’s call for design separation to enhance explainability, fault containment, and assurance traceability. 
	
	\textit{Lifecycle Oversight and Continuous Assurance:} 
	A key tenet of~\cite[Sec.~7]{ONR-TPR_2024} is lifecycle-spanning assurance, encompassing design-time verification, runtime monitoring, and post-deployment adaptation. Our model-driven pipeline (Section~3.3) operationalises this principle by maintaining traceability between PRISM analyses and GSN assurance cases~\cite[Sec.~7]{ONR-TPR_2024}. Placeholders, monitor bindings, and evidence cost tags are included to support runtime updates and evolution-time regeneration. Derived runtime monitors will observe environmental assumptions (e.g., radiation bounds), and assurance monitors will track probabilistic confidence over time. Violations trigger GSN goal reopening and selective re-verification, consistent with the emphasis on configuration management, drift detection, and adaptive retraining~\cite[Sec.~7]{ONR-TPR_2024}.
	
	\textit{Human--AI Interaction and Organisational Factors:}  
	~\cite[Sec.~5]{ONR-TPR_2024} stresses the critical role of human--AI interaction, advocating for explainable interfaces, intervention pathways, and traceable control transitions. In our study, handover events and override rules are explicitly modelled in RoboChart and verified to prevent unsafe operator actions during hazardous modes. The GSN model incorporates assumptions on manual intervention, and our verification ensures that no operator command can bypass \texttt{AIR\_SafetyWrapper} constraints in \emph{Caution} or \emph{Emergency Retrieval} modes.
	
	\textit{Documented, Dynamic Safety Cases:}  
	~\cite[Sec.~8]{ONR-TPR_2024} emphasises the importance of transparent and evolving assurance documentation. Our integration of formal verification artefacts into GSN structures ensures that assurance claims are both traceable and updatable. Design-time claims include placeholders for runtime validation and evolution-time updates, each annotated with effort cost and monitor identifiers. Violated assumptions or reduced confidence scores will trigger automatic reopening of affected GSN goals, with regeneration prioritised by cost and safety criticality.
	
	\section{Conclusion}
	\label{sec:conclusion}
	In this paper, we proposed a unified \emph{Continuous Assurance Framework} that integrates design-time, runtime, and evolution-time assurance activities into a model-driven, lifecycle-aware methodology. Through an illustrative case study involving an autonomous nuclear inspection robot, we instantiated the design-time stage of the framework using dual formal verification workflows: RoboChart with FDR4 for functional correctness, and PRISM for probabilistic risk quantification. These complementary analyses were integrated into structured GSN arguments enriched with placeholders and stereotypes to support traceability and automated updates. We discussed the framework in relation to the \emph{Trilateral AI Principles}, thus aligning with regulator-endorsed best practices.
	
	Future work will focus on extending the runtime and evolution-time phases of our framework. This includes implementing monitors to detect violations of design-time assumptions about sensor data or control loop behaviour, and updating assurance confidence scores based on real-time system data, using complementary techniques such as Bayesian belief networks, Dempster–Shafer evidence theory, and Baconian confidence models. We will investigate enhancing evolution-time tooling to support automated change impact analysis and selective regeneration of verification artefacts. 
	We also aim to broaden the framework’s applicability by incorporating learning-enabled components and adaptive behaviours, thereby addressing emerging challenges in \emph{assured autonomy} under epistemic uncertainty. 
	Another potential direction involves integration with an assurance case management platform to improve tool interoperability across the verification-to-certification pipeline.
	\section*{Acknowledgment}
	This work is supported by the \emph{Centre for Robotic Autonomy in Demanding and Long Lasting Environments} (CRADLE) under EPSRC grant EP/X02489X/1 and by the Royal Academy of Engineering under the \emph{Chairs in Emerging Technology} scheme.
	
	
	\balance
	\bibliographystyle{IEEEtran}
	\bibliography{Mybib}

\begin{thebibliography}{10}
\providecommand{\url}[1]{#1}
\csname url@samestyle\endcsname
\providecommand{\newblock}{\relax}
\providecommand{\bibinfo}[2]{#2}
\providecommand{\BIBentrySTDinterwordspacing}{\spaceskip=0pt\relax}
\providecommand{\BIBentryALTinterwordstretchfactor}{4}
\providecommand{\BIBentryALTinterwordspacing}{\spaceskip=\fontdimen2\font plus
\BIBentryALTinterwordstretchfactor\fontdimen3\font minus
  \fontdimen4\font\relax}
\providecommand{\BIBforeignlanguage}[2]{{%
\expandafter\ifx\csname l@#1\endcsname\relax
\typeout{** WARNING: IEEEtran.bst: No hyphenation pattern has been}%
\typeout{** loaded for the language `#1'. Using the pattern for}%
\typeout{** the default language instead.}%
\else
\language=\csname l@#1\endcsname
\fi
#2}}
\providecommand{\BIBdecl}{\relax}
\BIBdecl

\bibitem{Abeywickrama-CACM2024}
\BIBentryALTinterwordspacing
D.~B. Abeywickrama, A.~Bennaceur, G.~Chance, Y.~Demiris, A.~Kordoni, M.~Levine,
  L.~Moffat, L.~Moreau, M.~R. Mousavi, B.~Nuseibeh, S.~Ramamoorthy, J.~O.
  Ringert, J.~Wilson, S.~Windsor, and K.~Eder, ``On specifying for
  trustworthiness,'' \emph{Commun. ACM}, vol.~67, no.~1, p. 98–109, Dec.
  2023. [Online]. Available: \url{https://doi.org/10.1145/3624699}
\BIBentrySTDinterwordspacing

\bibitem{Rouff2022}
\BIBentryALTinterwordspacing
C.~Rouff and L.~Watkins, ``Assured autonomy survey,'' \emph{Foundations and
  Trends in Privacy and Security}, vol.~4, no.~1, pp. 1--116, 2022. [Online].
  Available: \url{http://dx.doi.org/10.1561/3300000027}
\BIBentrySTDinterwordspacing

\bibitem{Carlan2021}
C.~C{\^a}rlan, B.~Gallina, and L.~Soima, ``Safety case maintenance: A
  systematic literature review,'' in \emph{Computer Safety, Reliability, and
  Security}, I.~Habli, M.~Sujan, and F.~Bitsch, Eds.\hskip 1em plus 0.5em minus
  0.4em\relax Cham: Springer International Publishing, 2021, pp. 115--129.

\bibitem{Schleiss2022}
P.~Schleiss, F.~Carella, and I.~Kurzidem, ``Towards continuous safety assurance
  for autonomous systems,'' in \emph{2022 6th International Conference on
  System Reliability and Safety (ICSRS)}, 2022, pp. 457--462.

\bibitem{Hartsell2021b}
C.~Hartsell, S.~Ramakrishna, A.~Dubey, D.~Stojcsics, N.~Mahadevan, and
  G.~Karsai, ``Resonate: A runtime risk assessment framework for autonomous
  systems,'' in \emph{2021 International Symposium on Software Engineering for
  Adaptive and Self-Managing Systems (SEAMS)}, 2021, pp. 118--129.

\bibitem{Murphy2025}
L.~Murphy, M.~Carwehl, J.~P{\"a}{\ss}ler, and M.~Chechik, ``Towards systematic
  maintenance of assurance for evolving self-adaptive systems,'' in \emph{Proc.
  of the Engineering Reliable Autonomous Systems Conference (ERAS 2025)}.\hskip
  1em plus 0.5em minus 0.4em\relax IEEE, 2025.

\bibitem{Hoare1969}
\BIBentryALTinterwordspacing
C.~A.~R. Hoare, ``An axiomatic basis for computer programming,'' \emph{Commun.
  ACM}, vol.~12, no.~10, p. 576–580, Oct. 1969. [Online]. Available:
  \url{https://doi.org/10.1145/363235.363259}
\BIBentrySTDinterwordspacing

\bibitem{Miyazawa2019}
A.~Miyazawa, P.~Ribeiro, W.~Li, A.~Cavalcanti, J.~Timmis, and J.~Woodcock,
  ``\relax{RoboChart}: modelling and verification of the functional behaviour
  of robotic applications,'' \emph{Software and Systems Modeling}, vol.~18,
  no.~5, pp. 3097,3149, Jul. 2019.

\bibitem{PRISM}
\BIBentryALTinterwordspacing
\relax{PRISM Model Checker}, ``\relax{Property Manual},'' Online, 2024.
  [Online]. Available:
  \url{https://www.prismmodelchecker.org/manual/Main/AllOnOnePage}
\BIBentrySTDinterwordspacing

\bibitem{Belle2023}
A.~B. Belle, H.~Hemmati, and T.~C. Lethbridge, ``Position paper: A vision for
  the dynamic safety assurance of \relax{ML}-enabled autonomous driving
  systems,'' in \emph{2023 IEEE 31st International Requirements Engineering
  Conference Workshops (REW)}, 2023, pp. 297--301.

\bibitem{GSN}
\relax{Safety-Critical Systems Club}, ``\relax{Goal Structuring Notation},''
  \url{https://scsc.uk/gsn}, 2024.

\bibitem{ONR-TPR_2024}
{Canadian Nuclear Safety Commission}, {UK Office for Nuclear Regulation}, and
  {US Nuclear Regulatory Commission}, ``Considerations for developing
  artificial intelligence systems in nuclear applications,'' Canadian Nuclear
  Safety Commission; UK Office for Nuclear Regulation; US Nuclear Regulatory
  Commission, Trilateral Principles Report, September 2024.

\bibitem{Asaadi2020}
E.~Asaadi, E.~Denney, J.~Menzies, G.~J. Pai, and D.~Petroff, ``Dynamic
  assurance cases: A pathway to trusted autonomy,'' \emph{Computer}, vol.~53,
  no.~12, pp. 35--46, 2020.

\bibitem{BloomfieldRushby2020}
\BIBentryALTinterwordspacing
R.~Bloomfield and J.~Rushby, ``Assurance 2.0: A manifesto,'' \emph{arXiv
  preprint arXiv:2004.10474}, 2020. [Online]. Available:
  \url{https://arxiv.org/abs/2004.10474}
\BIBentrySTDinterwordspacing

\bibitem{Dong2023}
\BIBentryALTinterwordspacing
Y.~Dong, W.~Huang, V.~Bharti, V.~Cox, A.~Banks, S.~Wang, X.~Zhao, S.~Schewe,
  and X.~Huang, ``Reliability assessment and safety arguments for machine
  learning components in system assurance,'' \emph{ACM Trans. Embed. Comput.
  Syst.}, vol.~22, no.~3, Apr. 2023. [Online]. Available:
  \url{https://doi.org/10.1145/3570918}
\BIBentrySTDinterwordspacing

\bibitem{Bagheri2023}
M.~Bagheri, J.~Lamp, X.~Zhou, L.~Feng, and H.~Alemzadeh, ``Towards developing
  safety assurance cases for learning-enabled medical cyber-physical systems,''
  in \emph{Proceedings of the AAAI Workshop on Artificial Intelligence Safety
  (SafeAI)}, Washington, D.C., USA, Feb. 2023, february 13--14, 2023.

\bibitem{Avila2023}
R.~D. Avila and J.~B. Clark, ``Architecting systems for assured autonomy,'' in
  \emph{2023 IEEE International Conference on Assured Autonomy (ICAA)}, 2023,
  pp. 91--96.

\bibitem{Jahan2019}
S.~Jahan, M.~Pasco, R.~Gamble, P.~McKinley, and B.~Cheng, ``Mape-sac: A
  framework to dynamically manage security assurance cases,'' in \emph{2019
  IEEE 4th International Workshops on Foundations and Applications of Self*
  Systems (FAS*W)}, 2019, pp. 146--151.

\bibitem{Calinescu2018}
R.~Calinescu, M.~Češka, S.~Gerasimou, M.~Kwiatkowska, and N.~Paoletti,
  ``Efficient synthesis of robust models for stochastic systems,''
  \emph{Journal of Systems and Software}, vol. 143, pp. 140--158, 2018.

\bibitem{Denney2015ICSE}
E.~Denney, G.~Pai, and I.~Habli, ``Dynamic safety cases for through-life safety
  assurance,'' in \emph{2015 IEEE/ACM 37th IEEE International Conference on
  Software Engineering}, vol.~2, 2015, pp. 587--590.

\bibitem{Bourbouh2021}
\BIBentryALTinterwordspacing
H.~Bourbouh, M.~Farrell, A.~Mavridou, I.~Sljivo, G.~Brat, L.~A. Dennis, and
  M.~Fisher, ``Integrating formal verification and assurance: An inspection
  rover case study,'' in \emph{NASA Formal Methods: 13th International
  Symposium, NFM 2021, Virtual Event, May 24–28, 2021, Proceedings}.\hskip
  1em plus 0.5em minus 0.4em\relax Berlin, Heidelberg: Springer-Verlag, 2021,
  p. 53–71. [Online]. Available:
  \url{https://doi.org/10.1007/978-3-030-76384-8_4}
\BIBentrySTDinterwordspacing

\bibitem{Wei2024}
R.~Wei, Z.~Jiang, H.~Mei, K.~Barmpis, S.~Foster, T.~Kelly, and Y.~Zhuang,
  ``Automated model-based assurance case management using constrained natural
  language,'' \emph{IEEE Transactions on Computer-Aided Design of Integrated
  Circuits and Systems}, vol.~43, no.~1, pp. 291--304, 2024.

\bibitem{Denney2011}
\BIBentryALTinterwordspacing
E.~Denney, G.~Pai, and I.~Habli, ``Towards measurement of confidence in safety
  cases,'' in \emph{Proceedings of the 2011 International Symposium on
  Empirical Software Engineering and Measurement}, ser. ESEM '11.\hskip 1em
  plus 0.5em minus 0.4em\relax USA: IEEE Computer Society, 2011, p. 380–383.
  [Online]. Available: \url{https://doi.org/10.1109/ESEM.2011.53}
\BIBentrySTDinterwordspacing

\bibitem{WANG2019}
\BIBentryALTinterwordspacing
R.~Wang, J.~Guiochet, G.~Motet, and W.~Schön, ``Safety case confidence
  propagation based on dempster–shafer theory,'' \emph{International Journal
  of Approximate Reasoning}, vol. 107, pp. 46--64, 2019. [Online]. Available:
  \url{https://www.sciencedirect.com/science/article/pii/S0888613X18303505}
\BIBentrySTDinterwordspacing

\bibitem{Weinstock2013}
C.~B. Weinstock, J.~B. Goodenough, and A.~Z. Klein, ``Measuring assurance case
  confidence using baconian probabilities,'' in \emph{2013 1st International
  Workshop on Assurance Cases for Software-Intensive Systems (ASSURE)}, 2013,
  pp. 7--11.

\bibitem{Ferrando2021}
\BIBentryALTinterwordspacing
A.~Ferrando, L.~A. Dennis, R.~C. Cardoso, M.~Fisher, D.~Ancona, and
  V.~Mascardi, ``Toward a holistic approach to verification and validation of
  autonomous cognitive systems,'' \emph{ACM Trans. Softw. Eng. Methodol.},
  vol.~30, no.~4, May 2021. [Online]. Available:
  \url{https://doi.org/10.1145/3447246}
\BIBentrySTDinterwordspacing

\bibitem{Luckcuck2019}
\BIBentryALTinterwordspacing
M.~Luckcuck, M.~Farrell, L.~A. Dennis, C.~Dixon, and M.~Fisher, ``Formal
  specification and verification of autonomous robotic systems: A survey,''
  \emph{ACM Comput. Surv.}, vol.~52, no.~5, Sep. 2019. [Online]. Available:
  \url{https://doi.org/10.1145/3342355}
\BIBentrySTDinterwordspacing

\bibitem{Wei2024b}
\BIBentryALTinterwordspacing
W.~Li, P.~Ribeiro, A.~Miyazawa, R.~Redpath, A.~Cavalcanti, K.~Alden,
  J.~Woodcock, and J.~Timmis, ``Formal design, verification and implementation
  of robotic controller software via robochart and robotool,'' \emph{Auton.
  Robots}, vol.~48, no.~6, Jul. 2024. [Online]. Available:
  \url{https://doi.org/10.1007/s10514-024-10163-7}
\BIBentrySTDinterwordspacing

\bibitem{PRISM2GSN-GitHub}
D.~B. Abeywickrama, ``{PRISM2GSN: Eclipse Plugin for Transforming
  Specifications and Verification Results into GSN Argument Models},''
  \url{https://github.com/DhamindaA/prism2gsn-eclipse-plugin}, Aug. 2025.

\bibitem{Denney2023}
E.~Denney, R.~Lee, G.~J. Pai, and I.~Šljivo, ``\relax{QUASAR}: Quantifiable
  assurance cases for trusted autonomy,'' Air Force Research Laboratory
  Information Directorate, Tech. Rep. AFRL-RI-RS-TR-2023-162, 2023.

\bibitem{AdvoCATE-UG}
AdvoCATE, ``\relax{User Guide AdvoCATE (Version 1.10)},'' Jan. 2023.

\end{thebibliography}
	\onecolumn
	\newpage
	\appendix 
	
	\section*{A: PRISM Formal Model for the Case Study}
	
\begin{lstlisting}[
	caption={PRISM DTMC model created for case study scenario.},
	label={lst:prism},
	xleftmargin=0pt,        % flush with column edge
	framexleftmargin=0pt,   % frame flush too
	numbersep=6pt,          % space between numbers and code
	columns=fullflexible,   % tighter width; fewer overflows
	breaklines=true,
	breakatwhitespace=true,
	keepspaces=true,        % preserve alignment spaces
	tabsize=2,              % shrink tabs (default is wide)
	basicstyle=\ttfamily\scriptsize % smaller font if needed
	]
	// -------------------------------------------------------------
	// PRISM DTMC Model for Nuclear Radiation Inspection Robot with Safety-Wrapper
	// -------------------------------------------------------------
	
	dtmc
	// -------------------------------------------------------------
	// These constants define the environment and robot behavior parameters.
	
	// Radiation sampling probabilities (must sum to 1.0)
	const double p_rad_crit   = 0.02; // Probability of critical radiation (rad = 2)
	const double p_rad_med    = 0.08; // Probability of warning radiation (rad = 1)
	const double p_rad_safe   = 1 - p_rad_crit - p_rad_med; // Probability of safe radiation (rad = 0)
	
	// Collision and battery dynamics
	const double p_err        = 0.01; // Probability of entering a forbidden zone on any move
	const int    batt_dec     = 20;  // Battery drain per move in Patrol mode
	const int    batt_dec_cm  = 10;  // Battery drain per move in Caution mode (slower speed)
	const int    batt_threshold = 30; // Minimum battery level (SoC %) to safely continue mission
	
	// Radiation thresholds
	const int    R1 = 1; // Warning radiation threshold (triggers Caution mode)
	const int    R2 = 2; // Critical radiation threshold (triggers Emergency Retrieval mode)
	
	// -------------------------------------------------------------
	// Logical formulas used for guards and transitions
	// -------------------------------------------------------------
	formula is_warning  = (rad = 1); // true if Warning level radiation is sampled
	formula is_critical = (rad = 2); // true if Critical level radiation is sampled
	
	// -------------------------------------------------------------
	// Module: AIR_Navigator
	// Models the UGV's patrolling behavior
	// -------------------------------------------------------------
	module AIR_Navigator
	
	// State variables
	loc  : [0..6] init 0;     // Current location: 0-3 (waypoints), 4=Success, 5=Forbidden zone entry, 6=Abort
	batt : [0..100] init 100; // Battery level (state of charge), initialized to 100%
	rad  : [0..2] init 0;     // Most recent radiation sample: 0=Safe, 1=Warning, 2=Critical
	
	// ----------------------------------------------------------
	// Battery-abort transition: if battery is below threshold mid-mission
	// ----------------------------------------------------------
	[] loc < 4 & batt < batt_threshold ->
	(loc' = 6) & (batt' = batt) & (rad' = rad);
	
	// ----------------------------------------------------------
	// Movement in Patrol Mode (sw = 0)
	// Full speed, full battery drain, samples radiation at new location
	// ----------------------------------------------------------
	[] loc < 4 & batt >= batt_threshold & batt >= batt_dec & sw = 0 ->
	p_err :
	(loc' = 5) & (batt' = batt - batt_dec) & (rad' = rad) +
	(1 - p_err) * p_rad_crit :
	(loc' = loc + 1) & (batt' = batt - batt_dec) & (rad' = 2) +
	(1 - p_err) * p_rad_med :
	(loc' = loc + 1) & (batt' = batt - batt_dec) & (rad' = 1) +
	(1 - p_err) * p_rad_safe :
	(loc' = loc + 1) & (batt' = batt - batt_dec) & (rad' = 0);
	
	// ----------------------------------------------------------
	// Movement in Caution Mode (sw = 1)
	// Slower motion, reduced battery drain, continues route cautiously
	// ----------------------------------------------------------
	[] loc < 4 & batt >= batt_threshold & batt >= batt_dec_cm & sw = 1 ->
	p_err :
	(loc' = 5) & (batt' = batt - batt_dec_cm) & (rad' = rad) +
	(1 - p_err) * p_rad_crit :
	(loc' = loc + 1) & (batt' = batt - batt_dec_cm) & (rad' = 2) +
	(1 - p_err) * p_rad_med :
	(loc' = loc + 1) & (batt' = batt - batt_dec_cm) & (rad' = 1) +
	(1 - p_err) * p_rad_safe :
	(loc' = loc + 1) & (batt' = batt - batt_dec_cm) & (rad' = 0);
	
	// ----------------------------------------------------------
	// Emergency Retrieval Mode (sw = 2)
	// Halt in place, resets radiation to safe and no battery consumption 
	// ----------------------------------------------------------
	[] loc < 4 & sw = 2 & batt >= batt_threshold -> (loc' = loc) & (batt' = batt) & (rad' = rad);
	
	// ----------------------------------------------------------
	// Absorbing Terminal States
	// No transitions out of these; mission ends
	// ----------------------------------------------------------
	[] loc = 4 -> (loc' = 4); // Mission success
	[] loc = 5 -> (loc' = 5); // Forbidden zone entry
	[] loc = 6 -> (loc' = 6); // Abort (battery or retreat)
	endmodule
	
	// -------------------------------------------------------------
	// Module: AIR_SafetyWrapper
	// Supervisory controller for mode switching, velocity setting, and operator input control
	// -------------------------------------------------------------
	module AIR_SafetyWrapper
	
	sw       : [0..2] init 0;    // Mode: 0=Patrol, 1=Caution, 2=Emergency Retrieval
	vel      : [0..2] init 2;    // Velocity: 2=Full, 1=Slow, 0=Stopped
	op_used  : bool init false; // True if any operator command (hdng or vel) has been used
	
	// Transition: Patrol? Caution on warning-level radiation
	[] sw = 0 & is_warning ->
	(sw' = 1) & (vel' = 1);
	
	// Transition: Patrol/Caution? Emergency Retrieval on critical-level radiation
	[] sw <= 1 & is_critical ->
	(sw' = 2) & (vel' = 0);
	
	// Stay in Emergency Retrieval mode (absorbing)
	[] sw = 2 -> (sw' = 2) & (vel' = 0);
	
	// ----------------------------------------------------------
	// Operator/Navigator Commands
	// These commands are only enabled when sw = 0 (Patrol mode)
	// When triggered, they set op_used = true for later tracking
	// ----------------------------------------------------------
	[hdng] sw = 0 -> (sw' = sw) & (op_used' = true); // heading request
	[vel]  sw = 0 -> (sw' = sw) & (op_used' = true); // velocity command
	endmodule
	
	// -------------------------------------------------------------
	// Reward Structures
	// Used for quantitative analysis: cost, time, safety, and diagnostics
	// -------------------------------------------------------------
	// Reward: Each movement step forward (mission progress)
	rewards "moves"
	loc < 4 : 1;
	endrewards
	// Reward: Radiation exposure incidents (warning or critical)
	rewards "dose"
	rad >= 1 : 1;
	endrewards
	// Reward: Time spent in Caution mode (sw = 1)
	rewards "time_in_cm"
	sw = 1 : 1;
	endrewards
	// Reward: Time spent completely stopped (used for energy-saving diagnostics)
	rewards "time_stopped"
	vel = 0 : 1;
	endrewards
\end{lstlisting}
\end{document}